\documentclass[lettersize,journal]{IEEEtran}
\IEEEoverridecommandlockouts

\usepackage{graphicx}
\usepackage{amsmath}
\usepackage{amssymb}
\usepackage{booktabs}
\usepackage{caption}
\usepackage{makecell}
\usepackage{array}
\usepackage{multirow}
\usepackage{ragged2e}
\usepackage[pagebackref,breaklinks,colorlinks]{hyperref}
\usepackage{amsmath, xspace,  comment}
\usepackage{tikz}
\usepackage{orcidlink}
\usepackage{multirow}
\usepackage{enumitem}
\usepackage{algpseudocode}
\usepackage[linesnumbered]{algorithm2e}
\usepackage{soul}
\usepackage{colortbl}
\usepackage{balance}
\usepackage[most]{tcolorbox}
\tcbuselibrary{theorems}

\newtcbtheorem[]{observation}{\textbf{Observation}}%
{colback=white,colframe=white!65!black,fonttitle=\bfseries,coltitle=black}{}

\AtBeginDocument{%
  }

\newcommand{\etal}{{\em et al.}\xspace}
\newcommand{\eg}{{\em e.g.,}\xspace}
\newcommand{\ie}{{\em i.e.,}\xspace}

\newcommand{\BfPara}[1]{{\noindent\bf#1.}\xspace}
\newcommand{\ours}{{{AdViT${}$}\xspace}}

\newcommand*\cib[1]{\tikz[baseline=(char.base)]{ \node[shape=circle,fill=black,text=white,draw,inner sep=0.3pt] (char) {#1};}}

\setstcolor{red}

\hypersetup{
	colorlinks=true,
	urlcolor=blue!70!black,
	linkcolor=red!70!black,
	citecolor=purple,
} 

\setul{0.8pt}{}

\begin{document}

\title{\huge Breaking the Illusion of Security via Interpretation: Interpretable Vision Transformer Systems under Attack}

\author{Eldor~Abdukhamidov \orcidlink{0000-0001-8530-9477},
        Mohammed~Abuhamad~\orcidlink{0000-0002-3368-6024},
        Simon~S.~Woo~\orcidlink{0000-0002-8983-1542},\\
        Hyoungshick~Kim~\orcidlink{0000-0002-1605-3866},
        and~Tamer~Abuhmed~\orcidlink{0000-0001-9232-4843}
        
        \IEEEcompsocitemizethanks{\IEEEcompsocthanksitem Eldor Abdukhamidov, Hyoungshick~Kim,~Simon~S.~Woo~, and Tamer Abuhmed are with the Department of Computer Science and Engineering, Sungkyunkwan University, Suwon, South Korea.\protect
        (E-mail: abdukhamidov@skku.edu, swoo@skku.edu, hyoung@skku.edu, tamer@skku.edu). Mohammed Abuhamad is with the Department of Computer Science, Loyola University, Chicago, USA.\protect
        (E-mail: mabuhamad@luc.edu).\\ }  
        } 

\maketitle

\begin{abstract}
Vision transformer (ViT) models, when coupled with interpretation models, are regarded as secure and challenging to deceive, making them well-suited for security-critical domains such as medical applications, autonomous vehicles, drones, and robotics. However, successful attacks on these systems can lead to severe consequences. Recent research on threats targeting ViT models primarily focuses on generating the smallest adversarial perturbations that can deceive the models with high confidence, without considering their impact on model interpretations. Nevertheless, the use of interpretation models can effectively assist in detecting adversarial examples. This study investigates the vulnerability of transformer models to adversarial attacks, even when combined with interpretation models. We propose an attack called ``\ours{}'' that generates adversarial examples capable of misleading both a given transformer model and its coupled interpretation model. Through extensive experiments on various transformer models and two transformer-based interpreters, we demonstrate that \ours{} achieves a 100\% attack success rate in both white-box and black-box scenarios. In white-box scenarios, it reaches up to 98\% misclassification confidence, while in black-box scenarios, it reaches up to 76\% misclassification confidence. Remarkably, \ours{} consistently generates accurate interpretations in both scenarios, making the adversarial examples more difficult to detect. 
\end{abstract}


\section{Introduction}
Deep learning approaches have attained cutting-edge performance in various applications, and the field continues to expand. Recently, Vision Transformers (ViTs) have been introduced as a new technique that classifies data by dividing it into spatially separated parts \cite{dosovitskiy2020image}. Generally, ViTs are considered more robust against adversarial attacks compared to Convolutional Neural Networks (CNNs) in image classification \cite{bhojanapalli2021understanding, shao2021adversarial}. Furthermore, ViT-based systems become even more robust when coupled with an interpretation model \cite{10359476}. 

\begin{figure}[t]
    \centering
    \captionsetup{justification=justified}
    \includegraphics[width=0.9\linewidth]{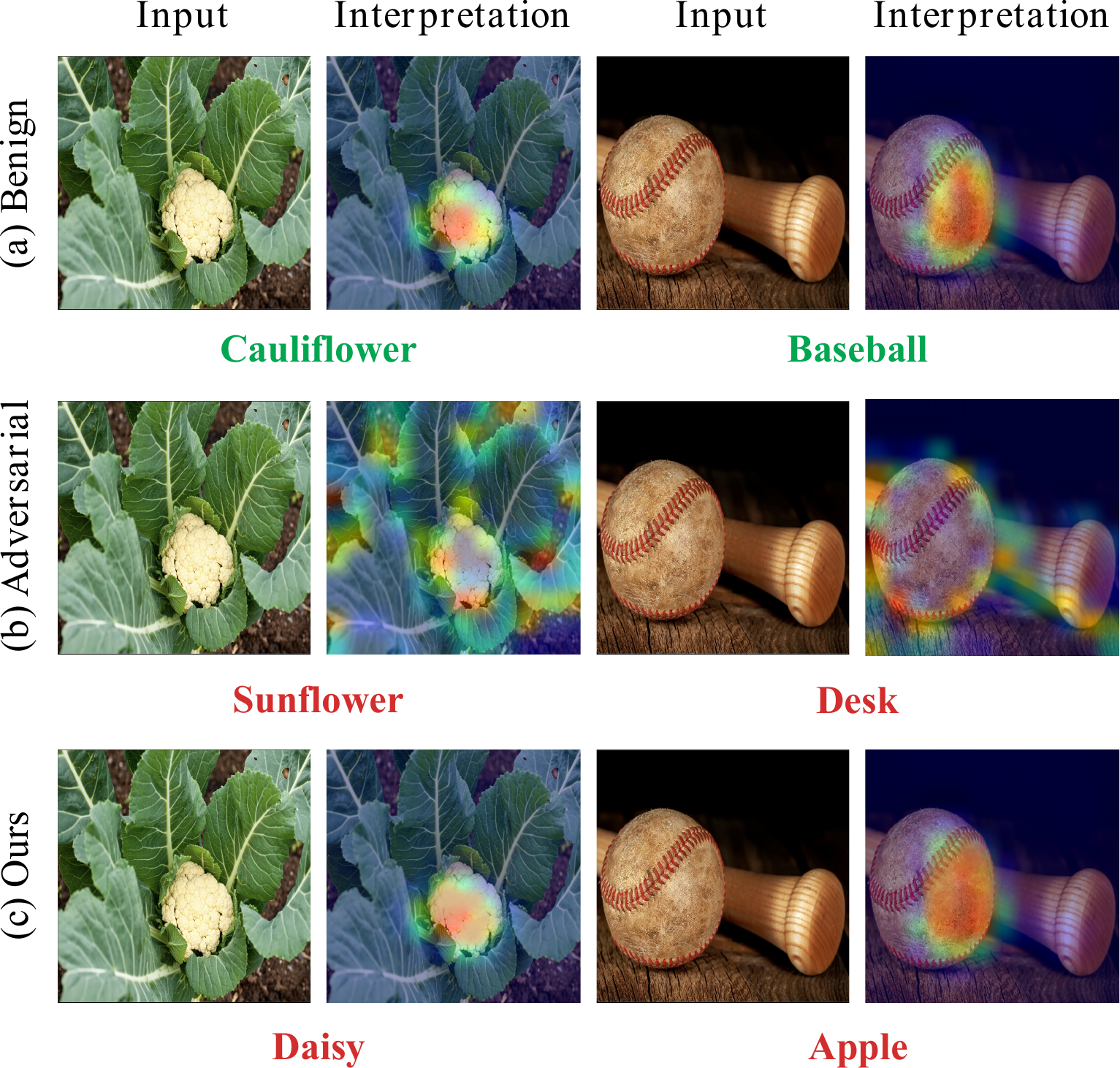}
    \caption{Example images comparing (a) benign samples, (b) samples subject to a regular adversarial attack, and (c) samples subject to our proposed attack, along with their  interpretations.}
    \label{fig:intro_images}
    \vspace{-1em}
\end{figure}

Adding interpretability as an integral component of machine learning pipelines improves their design, implementation, and adaptation by helping to detect and correct biases in the training dataset and identifying potential adversarial examples that could affect the final predictions. Moreover, interpretability ensures that only contextually relevant information is used for prediction. For example, \autoref{fig:intro_images} shows examples of a regular adversarial attack and the corresponding interpretation.

Until recently, it was believed that AI systems are more trustworthy and safe when integrated with interpretability methods and human involvement \cite{arrieta2020explainable}. However, the image classification field has recently shown that explainable methods are vulnerable and potential targets for malicious manipulations \cite{zhang2020interpretable, abdukhamidov2021advedge}. In \cite{ghorbani2019interpretation}, the study demonstrated that post-hoc methods are ineffective, resulting in considerable interpretation changes when a small amount of perturbation is applied to input samples. The study showed that two perceptually indistinguishable input images with the same predicted label and a small amount of perturbation could have significantly different interpretations. This is easily applicable to feature-importance interpretation methods, such as saliency maps, where the highlighted important pixels influence the model's decision. 
The study showed that slightly perturbed samples could have considerably different interpretations.

As more systematic studies have been conducted on the security of CNN models, little is known regarding interpretable deep learning systems (IDLSes) that employ transformer-based models. Recent studies have shown that transformer models are much more robust than CNNs \cite{bhojanapalli2021understanding,shao2021adversarial}. In this paper, we examine the security of various ViTs when coupled with interpretation models for the image classification task. We introduce a new {\underline{Ad}}versarial attack against {\underline{ViT}}s, {\ours}, that can generate adversarial samples that mislead the target transformer classifiers, such as DeiT-B, DeiT-S, DeiT-T, Swin-B, Swin-L, T2T-ViT-7, T2T-ViT-10, ViT-B, and ViT-L, and deceive their coupled interpretation models, including Transformer Interpreter and IA-RED$^2$, in both white-box and black-box settings. The key idea of \ours{} is to exploit vulnerabilities in the interaction between the transformers and their interpretation models. Unlike traditional ViTs adversarial attacks that only focus on fooling the classification model, \ours{} The attack strategically generates adversarial samples that mislead the target ViT classifiers and manipulate the output of their associated interpretation models. This dual deception is achieved through a novel attack technique that transforms the input image so that the changes are imperceptible to humans, making it even more challenging for detection methods that rely on interpretability as a security defense mechanism \cite{wang2020interpretability, noppel2023sok}. In this study, we show that \ours{} is applicable and efficient against models, such as ViT-B, Swin-T, MIT-B, and Vision-P, deployed in real-world scenarios, and is robust enough to circumvent various pre-processing defenses. 
We also discuss a possible interpretation-based method for detecting interpretation-guided adversarial samples using EfficientNet and a gradient-boosting classifier. 

\BfPara{Contributions} Our contributions are as follows:

\begin{itemize}[leftmargin=2ex]
    \item {\textbf{A Novel Joint Optimization Attack Framework}: We propose \ours{}, the first interpretation-guided adversarial attack framework specifically designed for transformer-based IDLSes. \ours{} introduces a new loss formulation that simultaneously optimizes for misclassification and interpretation similarity. The attack enables the generation of adversarial examples that fool the classification model and produce interpretations  highly similar to their benign inputs.}
    
    \item {\textbf{Mutation-Based Black-box Attack}: We extend \ours{} to black-box settings using a specialized mutation-based genetic algorithm (MGA). This approach significantly enhances transferability, allowing adversarial examples generated on a surrogate model to successfully deceive black-box ViT-based models and their interpreters, even when the attacker lacks complete model knowledge.}
    
    \item {\textbf{Comprehensive Evaluation}: We evaluate \ours{} on multiple transformer-based architectures (DeiT, Swin, T2T-ViT, ViT) and two transformer-based interpretation methods (Transformer Interpreter, IA-RED$^2$) in both white-box and black-box settings. The experiments show consistently high attack success rates, strong misclassification confidence, and closely-similar interpretation maps.}

    \item {\textbf{Robustness Against Real-World Models and Defenses}: We validate the practicality of \ours{} by successfully attacking four ViT-based models deployed as real-world APIs. Moreover, we show that \ours{} remains highly effective against several common defenses, including pre-processing transformations and adversarial training. We propose an interpretation-based ensemble detection strategy, suggesting potential technical countermeasures to mitigate the threat posed by \ours{}.}
\end{itemize}

\BfPara{Organization} The remainder of the paper is organized as follows: \autoref{sec:background} introduces the notations, threat models, and targeted interpretation models; \autoref{sec:methods} describes the proposed \ours{} attack and its formulation; \autoref{sec:experiment} provides the experiments and results; \autoref{sec:related} covers the related work; and \autoref{sec:conc} concludes the paper.

\section{Background} \label{sec:background}
This section introduces the notations, concepts, and targeted interpretation models essential for analyzing and optimizing attacks against vision transformer-based IDLSes. 

\subsection{Notations} \label{subsec:notations}
This work focuses on targeting image transformer-based classification models via white-box and black-box attacks. A transformer model with $n$ number of transformer blocks is defined as $\mathcal{F} = (f_1 \circ f_2 \circ f_3 \circ ... f_n) \circ f_{cls}$, where $f_i$ is the $i$-th transformer block consisting of multi-head self-attention and feed-forward layers, while $f_{cls}$ is the classification head, including the final norm layer with MLP-head. The model receives a sample image divided into $m$ patches and produces the processed patches within self-attention layers. In terms of classification, the processed patches are submitted to the final classification head to generate the output. Each transformer block ($f_i$) within the model helps extract the important features of the patches, and the classification head projects and relates the processed patches to the classes.
In the paper, $\mathcal{F}$ and $\mathcal{F}'$ represent a white-box and black-box transformer classifier, respectively, such that $\mathcal{F}(x)= c \in \mathcal{C}$ where $x$ is the input and $c$ is its category from a set of categories $\mathcal{C}$.

For interpretability, we consider post-hoc interpretations as this type of interpreter does not require any modification of the model architecture or parameters.
An interpreter $\mathcal{G}(x; \mathcal{F})= m$ produces an attribution map ($m$) that shows the importance of features in the input ($x$) based on the output of $ \mathcal{F}$ (\ie the value the $i\text{-th}$ element in $m$ ($m[i]$) reflects the importance of the $i\text{-th}$ element in $x$ ($x[i]$)).

\BfPara{Threat Model: Adversarial Objectives} 
 
The main goal of the attack is to make the classifier misclassify an adversarial sample $\hat{x}$ and to make the interpreter generate a similar interpretation $\hat{m}$ with its benign interpretation $m$.
Specifically, the attack aims to generate $\hat{x}$ such that \cib{1} $\mathcal{F}(\hat{x}) \neq c$;
\cib{2} $\mathcal{G}(\hat{x};\mathcal{F})=\hat{m}$ s.t. $\hat{m} \cong m$; and
\cib{3} $\hat{x}$ and the benign version $x$ should be visually imperceptible.

\BfPara{Threat Model: Adversarial Capabilities} This work assumes both white-box and
black-box settings. In the white-box scenario, the adversary has complete access to the transformer model $\mathcal{F}$ and the interpreter $\mathcal{G}$. 
In the black-box scenario, the adversary has limited knowledge and access, \ie query access, to the model $\mathcal{F}'$ (\eg output of the model).

\subsection{Targeted ViT Interpretation Models}

This work considers two state-of-the-art interpreters: Transformer Interpreter \cite{chefer2021transformer} and IA-RED$^2$ \cite{pan2021ia}. These interpreters were chosen based on their unique approaches to model interpretation, which offer distinct advantages in terms of efficiency, interpretive depth, and the ability to identify potential vulnerabilities or biases in transformer models.

The Transformer Interpreter was selected for its comprehensive analysis of the model's decision-making process, which is achieved through the combination of Layer-wise Relevance Propagation (LRP) and Deep Taylor Decomposition (DTD) techniques. By providing both high-level and fine-grained interpretations, this method offers a detailed understanding of the model's behavior, making it an ideal choice for studying the impact of adversarial attacks on transformer models.

On the other hand, IA-RED$^2$ was chosen for its ability to enhance the interpretive depth while minimizing redundancy in the interpretation. By identifying and removing redundant features, this method provides a more concise and informative interpretation, which can be particularly useful in identifying potential vulnerabilities or biases in the model. This focus on the most relevant features makes IA-RED$^2$ a valuable tool for analyzing the robustness of transformer models against adversarial attacks.

By comparing the results obtained using both interpretation methods, we can gain in-depth understanding of the strengths and weaknesses of transformer models in the face of adversarial attacks and identify potential strategies for improving their robustness.

\BfPara{Transformer Interpreter~\cite{chefer2021transformer}} 
The Transformer Interpreter is based on the Deep Taylor Decomposition \cite{montavon2017explaining}, which propagates local relevance through layers, including skip connections and attention layers. It adopts LRP-based Propagation \cite{bach2015pixel} relevance to measure the importance scores of a given sample for every layer of the transformer model, combining all those scores by relevancy scores and class-specific gradients.

The mathematical foundation of this interpreter is articulated through the equation $ C = \bar{A}^{(1)} \cdot \bar{A}^{(2)} \cdot ... \cdot \bar{A}^{(i)}$, where $\bar{A}^{(i)}$  represents a modified attention map for a given block $i$. Each $\bar{A}^{(i)}$ is formulated as $\bar{A}^{(i)} = I + \mathbb{E}_h (\nabla A ^{(i)} \odot R^{(s_i)})^{\textbf{+}}$, encapsulating the attention coefficients for each token within the block. Here, $\mathbb{E}_h$ denotes the mean across the attention heads, $R^{(s_i)}$ the relevance score linked to the softmax operation's layer $s_i$, $\odot$ symbolizes the Hadamard operation for element-wise multiplication, and $^+$ signifies the rectification operation $max(0,a)$, effectively isolating positive contributions.
An important feature of this interpreter is how it deals with skip connections in transformer blocks. It uses an identity matrix $I$ for each token, which helps to prevent important details from getting lost as they move through different layers. This means that it can keep track of changes in the input as it goes through the transformer model, ensuring that nothing important is missed.

\BfPara{IA-RED$^2$~\cite{pan2021ia}} 
The main idea of the Interpretability-Aware Redundancy Reduction (IA-RED$^2$) interpreter is derived from the architecture of the multi-head self-attention layer (MSA), called the multi-head interpreter, to evaluate whether a given patch token is important. The model is divided into several groups, each containing MSA and feed-forward network (FFN) blocks and one multi-head interpreter. 
The multi-head interpreter evaluates the input before passing it to the blocks inside each group to calculate the informative score $I_{ij}$, where $i$ and $j$ are the positions of the input token and the group, respectively. If the score ($I_{ij}$) is below the threshold (\eg 0.5), the patch $x_{i}$ is considered uninformative and will be ignored in the following groups.
The score ($I_{ij}$) is calculated as:
$$
    {I_{ij}} = \frac{1}{H} \sum_{h} \phi (F_{q}^{h}(x_{i}) \ast F_{k}^{h}(P_{j})),
$$
where $P_{j}$ is the policy token in the $j$-th multi-head interpreter to estimate the importance of the input token $x_{i}$, $H$ is the number of heads, $F_{q}^{h}$ and $F_{k}^{h}$ are the linear layers of the $h$-th head for the patch and policy token, $\ast$ is the dot product, $\phi$ is the sigmoid activation function. The subscript q stands for `query,' a term borrowed from the transformer architecture, where each input element is transformed into a query vector. The subscript k, \ie `key,' is another concept from transformer models, where elements are also transformed into key vectors for comparison against queries.

The reinforcement method is used to optimize the interpreter to make it more efficient and accurate.
The framework optimizes each multi-head interpreter using the expected gradient as follows:
$$
    \nabla_{W_j} J = E_{u \sim \pi}\left[ A \nabla_{W_j} \sum_{i=0}^N \log \left[I_{ij} u_i + (1 - I_{ij})(1 - u_i)\right] \right],
$$
where $E_{u \sim \pi}$ is the expected reward for the gradient computation, $A$ is the reward score, $W_j$ is the representation of the parameters in the $j$-{th} multi-head interpreter, $I_{ij}$ is the informative score of the $i$-{th} input token in the $j$-th multi-head interpreter, and $u_i$ is the configuration parameter accepting only binary values, in which 0 means ignoring the token.

\section{Overview of \ours{}} \label{sec:methods}
 
This section describes the \ours{} attack strategy to create dual-objective adversarial examples capable of deceiving both a targeted transformer model and its interpretation model for white- and black-box scenarios.

\begin{figure}[th]
    \centering
    \captionsetup{justification=justified}
    \includegraphics[width=0.9\linewidth]{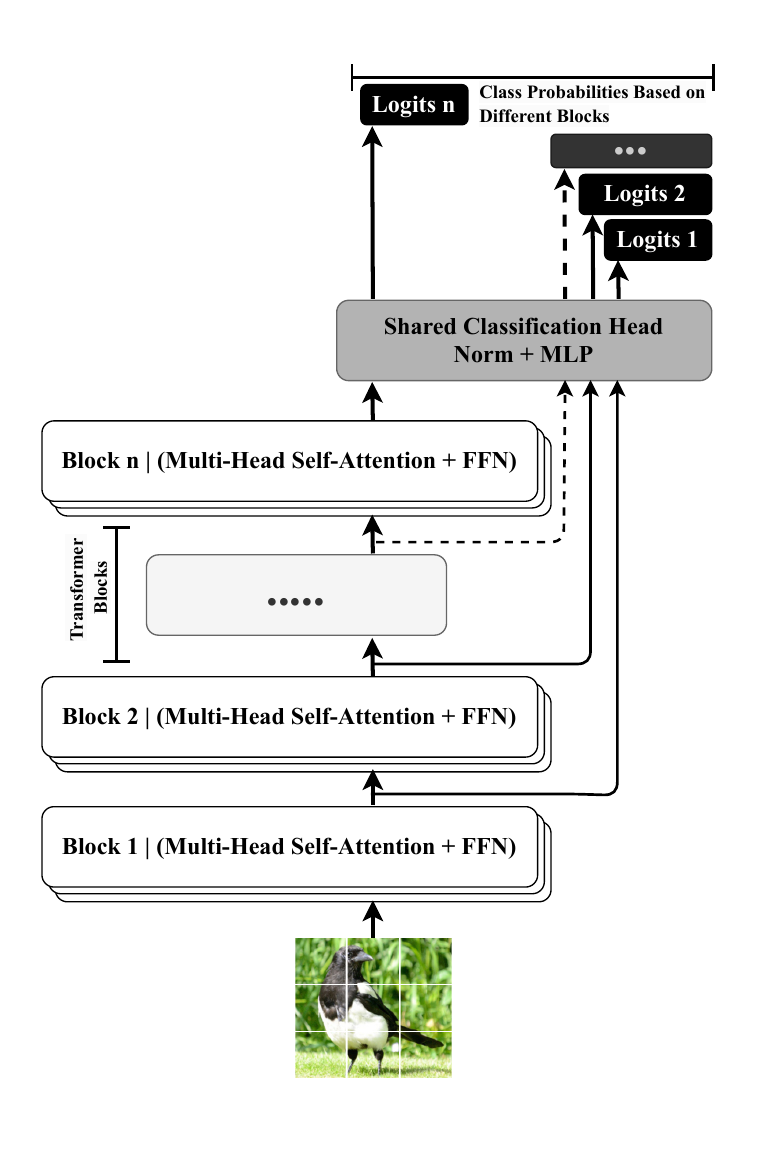}
    \caption{Modified transformer architecture to extract discriminative information of a given input from all blocks to enable \ours{} adversarial attack. Unlike traditional transformer models where only the final block (\(Block_n\)) connects to the classification layer, 
    this design maps each intermediate block ${Block_i}$ to a shared classification head (Norm + MLP). By obtaining separate logits from each block's output, we can leverage richer block-level information for more powerful adversarial optimization—without retraining or altering the Transformer blocks themselves.}
    
    \label{fig:vit_architecture}
    \vspace{-1em}
\end{figure}
\subsection{\ours{}: Attack Formulation} \label{sec:attack_formulation}

\BfPara{Leveraging Block-specific Features} Creating effective perturbations across all input patches requires identifying the features within each transformer block \(\{f_{i}\}_{i=1}^{n}\) that are most relevant to the target class. 
Our approach is inspired by the transferability logic of the ATViT attack~\cite{naseer2022on}, but unlike ATViT, which requires training or adapting the classification layer, we do not modify or retrain the underlying transformer weights.
{Instead, we introduce a mechanism that shares a single classification head \(g\) with each transformer block \(f_{i}\). 
Formally, define:
\begin{align*}
    \mathcal{F} &= \{f_{i}\}_{i=1}^{n} \;\circ\; g  \quad\text{where,}\; \;
    \mathcal{F}_1 = f_{1} \;\circ\; g, \quad\text{and}\\
    \mathcal{F}_{j} \;&=\; \Bigl(f_{1} \circ \dots \circ f_{j}\Bigr) \;\circ\; g
    \quad\text{for } j =  \{2, \ldots, n\},
\end{align*}
where: \(f_{i}\) is the \(i\)-th transformer block (including multi-head self-attention and feed-forward layers),
\(g\) is the shared classification head (\eg a normalization layer plus an MLP),
\(\circ\) denotes function composition of transformer blocks and the classification head.

In this formulation, \(\mathcal{F}_{j}\) represents a partial model that processes the input through the first \(j\) blocks \(f_1, f_2, \dots, f_j\), and then applies \(g\) to produce logits. By connecting each intermediate block \(f_j\) to the classification head \(g\) (instead of solely the last block \(f_n\)), we can extract critical \emph{layer-specific} discriminative information for adversarial attack purposes, all without retraining the transformer blocks themselves.}

\autoref{fig:vit_architecture} shows the architecture of the modified model for the attack. This architecture is used to extract discriminative information across patches based on the classification layer of each transformer block. 
This is done by passing patches in each transformer block and updating them to retain important information and discarding irrelevant information. The final prediction for the updated patches is then used to calculate the adversarial loss function. As the network updates the patches in each transformer block, it distills crucial information, which allows the final prediction to retain the most discriminative information. This is a crucial step in our proposed method, as it allows us to compute the loss of input more effectively. The attack generates adversarial samples that consider classification and interpretation models using extracted information.

We note that while our current approach does not alter the existing weights, we expect that fine-tuning or retraining the model could further enhance attack transferability or interpretation quality.

\BfPara{Attack Formulation} The traditional attack formulation to generate perturbation is as follows:
\begin{equation} \label{eq:finalFormula}
\begin{split}
    \hat{x}^{(i+1)} =  \amalg _{\mathcal{B}_{\varepsilon}(x)}\left(\hat{x}^{(i)} - \alpha .~ sign(\nabla_{\hat{x}}\ell_{adv}(\hat{x}^{(i)}))\right),
\end{split}
\end{equation}
where $ \amalg$, $\alpha$, and $\mathcal{B}_{\varepsilon}(x)$ represent the projection operator, learning rate, and the norm ball respectively. 
The symbol $\varepsilon$ in the attack formulation represents the size of the allowable perturbation and $\ell_{adv}$ is the overall adversarial loss. Using \autoref{eq:finalFormula} to generate adversarial samples based only on classification loss is not effective, as they can be easily detected by interpretation models (see \autoref{fig:intro_images}). 
In such cases, it is necessary to optimize the attack to generate perturbations that can mislead the interpretation models. To produce stealthy adversarial samples that fool the classification model and interpreter, we minimize the overall adversarial loss in terms of both classification loss $\ell_{cls}$ and interpretation loss $\ell_{int}$: 
\begin{equation} \label{eq:mainFormula}
    \ell_{adv} = \min_{\hat{x}} \quad \ell_{cls}(\mathcal{F}(\hat{x})) + \lambda ~\ell_{int}(\mathcal{G}(\hat{x};\mathcal{F}), m)
\end{equation}

Here, $\ell_{cls}$ is the cross entropy classification loss, $\ell_{int}$ is the difference of adversarial
interpretation maps and benign interpretation maps, and the hyper-parameter $\lambda$ balances $\ell_{cls}$ and $\ell_{int}$. 

We define $\ell_{cls}$ as:
\begin{equation} \label{eq:detailedClsFormula}
    \ell_{cls} = -\frac{1}{n} \sum_{j=1}^{n} \log\left(\frac{e^{g_{c}(\mathcal{F}_j(\hat{x}))}}{\sum_{k=1}^{|\mathcal{C}|} e^{g_{k}(\mathcal{F}_j(\hat{x}))}}\right)
\end{equation}

Here, $\mathcal{F}j(\hat{x})$ denotes the output of the $j$-th transformer block processed by the adversarial sample $\hat{x}$, and $g_{c}(\mathcal{F}_j(\hat{x}))$ represents the logit for the true class $c$, produced by the classification head based on the $j$-th block's output. The function aims to maximize the misclassification at all levels of the transformer model.

We define $\ell_{int}$ as:
\begin{equation} \label{eq:detailedIntFormula}
    \ell_{int} = \sum_{i=1}^{m} w_i \cdot (\mathcal{G}(\hat{x}; \mathcal{F})_i - m_i)^2
\end{equation}

Here, $m_i$ and $\mathcal{G}(\hat{x}; \mathcal{F})_i$ represent the importance scores of the $i$-th feature (or patch) in the benign and adversarial interpretation maps, respectively. The weight $w_i$ could be introduced to prioritize features based on their relevance to the classification decision, although for simplicity, it could initially be set to 1 for all features, which implies equal importance. This formulation computes a weighted sum of squared differences across all features, encouraging the adversarial sample to maintain a similar interpretation to the benign sample.

The primary objective of \ours, as encapsulated in \autoref{eq:mainFormula}, is to minimize the combined loss function that integrates the classification and interpretation losses of the adversarial input $\hat{x}$. 
This approach is optimized using the revised loss functions of $\ell_{cls}$ and $\ell_{int}$, as shown in \autoref{eq:detailedClsFormula} and \autoref{eq:detailedIntFormula}. 
These formulations leverage the inherent structure of transformer blocks to enhance the attack's effectiveness, making it applicable to a wide range of transformer-based models. 
By emphasizing the need for misclassification through $\ell_{cls}$ while simultaneously ensuring stealth through $\ell_{int}$, our method achieves a sophisticated balance, with the aim of minimizing the combined loss.

\subsection{\ours{} Optimization for Black-box Settings}
Investigating the applicability of \ours in a black-box scenario, we employ a modified mutation-based MGA \cite{harvey2009microbial} to generate adversarial examples against black-box models. 
Using adversarial samples generated against a white-box model $\mathcal{F}$ as the initial population, the MGA evolves the population to generate an adversarial sample that can fool the black-box model $\mathcal{F}'$ and its interpreter $\mathcal{G}'$. \ours{} in the black-box scenario is described in \autoref{alg:algorithm_mga}. 
The attack consists of genetic algorithm operators: \textit{initialization} (line 1-2), \textit{selection} (line 4-6), \textit{crossover} (line 7), \textit{mutation} (line 8), and \textit{population update} (line 12).

\RestyleAlgo{ruled}
\begin{algorithm}[t]
\caption{\small \ours{} attack in black-box settings}\label{alg:algorithm_mga}
\KwData{Source model $\mathcal{F}$, interpreter $\mathcal{G}$, input $x$, original category $c$, perturbation threshold $\epsilon$, mutation rate $mr$, crossover rate $cr$,  target model $\mathcal{F}'$, population size $n$, generation $G$,} 
\KwResult{Adversarial sample $\hat{x}$}
$x'$ = our\_attack($\mathcal{F}$, $\mathcal{G}$, $x$, $n$) \\
$pop$ = init\_population($x$, $x'$, $\epsilon$) \\
\For{$g\leftarrow 1$ \KwTo $G$}{
$p_1$, $p_2$ = random\_select($pop$)\\
$v_1$, $v_2$ = get\_fitness($\mathcal{F}'$, $x$, $p_1$, $p_2$)\\
$loser$, $winner$ = sort\_by\_fitness($p_1$, $p_2$, $v_1$, $v_2$)\\
$child$ = crossover($cr$, $loser$, $winner$)\\
$child$ = mutation($mr$, $child$)\\
\If {$f(child) \neq c$}{
return $child$
}
$pop$ = update\_population($pop$, $child$)
}
\end{algorithm}

\textbf{Initialization:} 
We generate adversarial samples by \ours{} in white-box settings and provide them as the initial population for MGA (\ie $\Psi: ~\{\psi_1, \psi_2, \dots, \psi_n\} $, where $n$ is the size of the population). 
We selected five as the population size, which provides the best trade-off between attack effectiveness and time complexity, based on our experiments. 

\textbf{Fitness function:} The function evaluates the quality of the samples in the population and helps to improve their evolution toward the optimal population. 
In our attack, we evaluate each individual in the population by applying a loss function (\ie relative cross entropy) as the fitness function, which is based on the classification confidence and perturbation size \cite{chen2019poba}. 
Loss values reflect the fitness scores of the samples in the population where a higher fitness score is desired to achieve the attack.
If a sample from the initial population is successful, the attack ends as the criterion is met (\eg when the transferability of white-box attacks is high).

\textbf{Selection:} Samples with high fitness scores have a higher chance of passing along their features to the next generation. \cite{back1996evolutionary}. 
To maintain diversity and high interpretability (passed from the initial generation), we randomly select two samples from a population, one (winner) with a high fitness score and another (loser) with a relatively lower score, to pass on to the crossover phase.

\textbf{Crossover:} The new offspring (adversarial sample) is generated by transferring the genetic data of the winner $\psi_{winner}$ and the loser $\psi_{loser}$ with the predefined crossover rate $cr$: $\psi_{child} = \psi_{winner} \ast S_{cr} + \psi_{loser} \ast (1 - S_{cr})$, where $S_{cr}$ is a mask matrix with the values of 1 and 0. $S_{cr} = 1 \quad rand(0, 1) < cr ~ 
otherwise ~0$, where $rand(0, 1)$ generates uniformly distributed numbers between 0 and 1. 
In the experiment, we use a crossover rate of $0.7$.

\textbf{Mutation:} The process further diversifies the population through another binary mask: $\psi_{child} = -\psi_{child} \ast S_{mr} + \psi_{child} \ast (1 - S_{mr})$, where $S_{mr}$ is a mask matrix based on the mutation rate $mr$.

We use a mutation rate of $1e\text{-}4$.

\textbf{Population update:} For continuous evolution, the population is updated by keeping the winners and replacing the losers with the new generation (\ie the mutated offspring).

The proposed modified genetic algorithm differs from the traditional ones by factoring in both interpretability and output classification when creating offspring. Specifically, we found that traditional GA-generated adversarial examples often fail to fool interpreter models, as they only consider children with high fitness scores (\ie based on classification output), resulting in a lack of diversity among future generations. Our proposed approach uses a strategy that considers children with low fitness scores (\ie losers) to generate adversarial examples to promote a higher degree of diversity and better control over the interpretation. This allows for a more effective attack on interpretability in addition to classification.

\begin{table}
\centering
\caption{Parameter configuration for the attack using perturbation generation (PGD) and genetic algorithm.}
\label{tab:parameters}
\resizebox{0.9\linewidth}{!}{%
\begin{tabular}{ccc} 
\toprule
\textbf{Algorithm}                                & \textbf{Parameter}         & \textbf{Values}  \\ 
\midrule
\multirow{4}{*}{\textbf{Perturbation generation}} & \# iterations              & 20               \\ 

                                                  & $\ell_{adv}$ coefficient ($\lambda$)       & 10               \\ 
                                                  & max. search step size $\alpha_{max}$ & 0.08             \\ 
                                                  & Perturbation threshold ($\epsilon$)     & 0.031            \\ 
\midrule
\multirow{3}{*}{\textbf{Genetic Algorithm}}       & Mutation rate ($mr$)              & 1e-4             \\ 
                                                  & Crossover rate ($cr$)            & 0.7              \\ 
                                                  & Population size ($n$)           & 5                \\
\bottomrule
\end{tabular}
}\vspace{-1em}
\end{table}

\section{Experiments} \label{sec:experiment}

We evaluate \ours{} attack on different vision transformer models and interpreters. Our analysis aims to answer the following questions:
{\cib{1} \emph{How effective is it to attack vision transformer models and their coupled interpreters?}
\cib{2} \emph{Are the adversarial examples transferable across various vision transformer models?}
\cib{3} \emph{Is it practical to attack the vision transformer models in black-box settings?}
\cib{4} \emph{Is it possible to attack real-world vision transformer models? }}
For the reproducibility of our experiments, our code, data, and models are available at ({\em\url{}\ul{https://github.com/InfoLab-SKKU/AdViT}}). For comparison, existing attacks were implemented under our experimental environment and evaluated on the same dataset.

\subsection{Experimental Settings}
\BfPara{Datasets} 
For our experiments, we use 1,000 images from the validation set of ImageNet that are classified correctly with a confidence higher than 70\% by the target ViT model. 

\BfPara{ViT Models}
In white-box settings, we target DeiT-B, DeiT-S, DeiT-T \cite{touvron2021training}, {Swin-B, Swin-L \cite{liu2021swin}, T2T-ViT-7, T2T-ViT-10 \cite{Yuan_2021_ICCV}, ViT-B and ViT-L \cite{dosovitskiy2020image}} models. 
In black-box settings, we use the same white-box models as surrogate models to attack ViT family models (ViT-B and ViT-L) \cite{dosovitskiy2020image}. 
In the realistic black-box scenario, 
we also demonstrate the effectiveness of \ours{} against real-world APIs of four ViT models: ViT-B by Google \cite{dosovitskiy2020image}, SWIN-T by Microsoft \cite{liu2021swin}, MIT-B3 by Nvidia \cite{DBLP:journals/corr/abs-2105-15203}, and Vision-perceiver-learned by DeepMind \cite{DBLP:journals/corr/abs-2107-14795}.  

\BfPara{Interpreters}
We employ two interpreters: Transformer Interpreter \cite{chefer2021transformer} and IA-RED$^2$ \cite{pan2021ia}. Those interpreters utilize different characteristics of the model to generate interpretations.

\BfPara{Metrics} 
The evaluation metrics are divided into classification and interpretation metrics. Classification metrics include {attack success rate} and {misclassification confidence} (\ie adversarial confidence).

Metrics used for interpreters are qualitative comparison and IoU score.

In addition, we adopt another metric (\textit{noise rate}) to measure the perturbation size.
The description of each metric is as follows.

\begin{itemize}[leftmargin=*]
    \item \ul{Attack success rate}: It calculates the ratio of successful attack cases to the total attack cases.
    \item \ul{Misclassification confidence}: We measure the probability (confidence score) of an adversarial sample assigned by the target model. We calculate the average probability of adversarial samples being successfully misclassified.
    \item \ul{Qualitative comparison}: This method is used to verify whether the interpretation results are perceptually similar. Every interpretation map is manually checked to see if it is identical to its benign interpretation map or if the interpretation is reliable.
    \item \ul{IoU score} (Intersection-over-Union):  This metric is used to quantify the similarity between two arbitrary shapes. It encodes the shape properties of interpretation maps, \eg \textit{height, width, and location} into region properties and calculates the intersection areas between the predictions and the ground truths. It is widely employed to evaluate object detection, segmentation, and tracking: 
    $$\text{IoU}(m) = | O(m) \bigcap O(m_{\circ})| ~ / ~| O(m) \bigcup  O(m_{\circ}) |,$$ where, $m$ is the attribution map of samples when the universal perturbation is added and $m_{\circ}$ is the attribution map of samples without any perturbation. In our case, we compare an adversarial interpretation map with the benign interpretation map based on (shapes, positions, and areas), for which the metric can be applied.
    \item \ul{Noise rate}: Perturbation amount is calculated using the structural similarity index (SSIM) \cite{wang2004image}. SSIM measures the similarity score, and we find the non-similarity portion using that score (\ie $\text{noise\_rate} = 1 - \text{SSIM}$). 
\end{itemize}

{The table containing the parameter values for the experimental settings is presented in  \autoref{tab:parameters}.}

\begin{table*}[tb]
\centering

\caption{{White-box scenario: Comparison of misclassification confidence against various ViT-based models and two interpreters (Transformer Interpreter and IA-RED$^2$). The gray-shaded columns (PGD and ATViT) represent non-interpreter baseline attacks, while other methods (ADV$^2$, AdvEdge, and \ours{}) use interpretative information in crafting perturbations.}}
\label{tab:mc}
\resizebox{0.9\linewidth}{!}{%
\begin{tabular}{c|cc|ccc|ccc} 
\toprule
\multirow{2}{*}{} & \multicolumn{2}{c|}{} & \multicolumn{3}{c|}{\textbf{Transformer Interpreter}} & \multicolumn{3}{c}{\textbf{IA-RED$^2$}}             \\ 
\cline{2-9}
                  & \cellcolor[rgb]{0.753,0.753,0.753} \textbf{PGD} &  \cellcolor[rgb]{0.753,0.753,0.753} \textbf{ATViT} & \textbf{ADV$^2$} & \textbf{AdvEdge} & \textbf{Ours (\ours{})}  &  \textbf{ADV$^2$} & \textbf{AdvEdge} & \textbf{Ours (\ours{})}  \\ 
 \cmidrule{2-9}
\textbf{DeiT-B}  & \cellcolor[rgb]{0.753,0.753,0.753} 0.59$\pm$0.20  & \cellcolor[rgb]{0.753,0.753,0.753} 0.49$\pm$0.21 & 0.62$\pm$0.22          & 0.64$\pm$0.22             & \textbf{0.78$\pm$0.18}   & 0.19$\pm$0.24          & 0.21$\pm$0.24             & \textbf{0.45$\pm$0.20}  \\ 
 
\textbf{DeiT-S}   & \cellcolor[rgb]{0.753,0.753,0.753} 0.58$\pm$0.19  & \cellcolor[rgb]{0.753,0.753,0.753} 0.51$\pm$0.23 & 0.63$\pm$0.22          & 0.64$\pm$0.22             & \textbf{0.65$\pm$0.19}  & 0.17$\pm$0.22          & 0.17$\pm$0.23             & \textbf{0.36$\pm$0.20}  \\ 
 
\textbf{DeiT-T}   & \cellcolor[rgb]{0.753,0.753,0.753} 0.56$\pm$0.19  & \cellcolor[rgb]{0.753,0.753,0.753} 0.50$\pm$0.22 & 0.61$\pm$0.22          & 0.62$\pm$0.22             & \textbf{0.72$\pm$0.18}   & 0.18$\pm$0.21          & 0.20$\pm$0.21             & \textbf{0.43$\pm$0.19}  \\
 
\textbf{Swin-B}   & \cellcolor[rgb]{0.753,0.753,0.753} 0.60$\pm$0.21  & \cellcolor[rgb]{0.753,0.753,0.753} 0.53$\pm$0.23 &     0.60$\pm$0.20      &      0.60$\pm$0.20        & \textbf{0.95$\pm$0.10}   &   0.25$\pm$0.20      &        0.25$\pm$0.20      & \textbf{0.59$\pm$0.12}  \\
 
\textbf{Swin-L}   & \cellcolor[rgb]{0.753,0.753,0.753} 0.59$\pm$0.22  & \cellcolor[rgb]{0.753,0.753,0.753} 0.49$\pm$0.24 &     0.58$\pm$0.21      &        0.60$\pm$0.21      & \textbf{0.98$\pm$0.08}   &      0.29$\pm$0.21    &         0.31$\pm$0.21     & \textbf{0.62$\pm$0.10}  \\
 
\textbf{T2T-ViT-7}  & \cellcolor[rgb]{0.753,0.753,0.753} 0.58$\pm$0.15  & \cellcolor[rgb]{0.753,0.753,0.753} 0.55$\pm$0.19 &      0.59$\pm$0.17     &         0.61$\pm$0.18     & \textbf{0.92$\pm$0.11}   &     0.26$\pm$0.18      &       0.26$\pm$0.18       & \textbf{0.60$\pm$0.14}  \\
 
\textbf{T2T-ViT-10}   & \cellcolor[rgb]{0.753,0.753,0.753} 0.56$\pm$0.18  & \cellcolor[rgb]{0.753,0.753,0.753} 0.54$\pm$0.20 &      0.55$\pm$0.19     &        0.59$\pm$0.19      & \textbf{0.94$\pm$0.16}    &     0.20$\pm$0.19      &        0.23$\pm$0.19      & \textbf{0.59$\pm$0.17}  \\
 
\textbf{ViT-B}   & \cellcolor[rgb]{0.753,0.753,0.753} 0.55$\pm$0.19  & \cellcolor[rgb]{0.753,0.753,0.753} 0.56$\pm$0.15 &      0.60$\pm$0.13     &         0.60$\pm$0.13     & \textbf{0.97$\pm$0.04}   &   0.38$\pm$0.14      &       0.42$\pm$0.14       & \textbf{0.95$\pm$0.13}  \\
 
\textbf{ViT-L}   & \cellcolor[rgb]{0.753,0.753,0.753} 0.55$\pm$0.20  & \cellcolor[rgb]{0.753,0.753,0.753} 0.54$\pm$0.16 &      0.61$\pm$0.12     &        0.58$\pm$0.12      & \textbf{0.96$\pm$0.05}   &     0.40$\pm$0.13      &        0.43$\pm$0.13      & \textbf{0.93$\pm$0.05}  \\
 \bottomrule
\end{tabular}
}

\end{table*}

\begin{figure*}
    \centering
    \captionsetup{justification=justified}
    \includegraphics[width=0.95\linewidth]{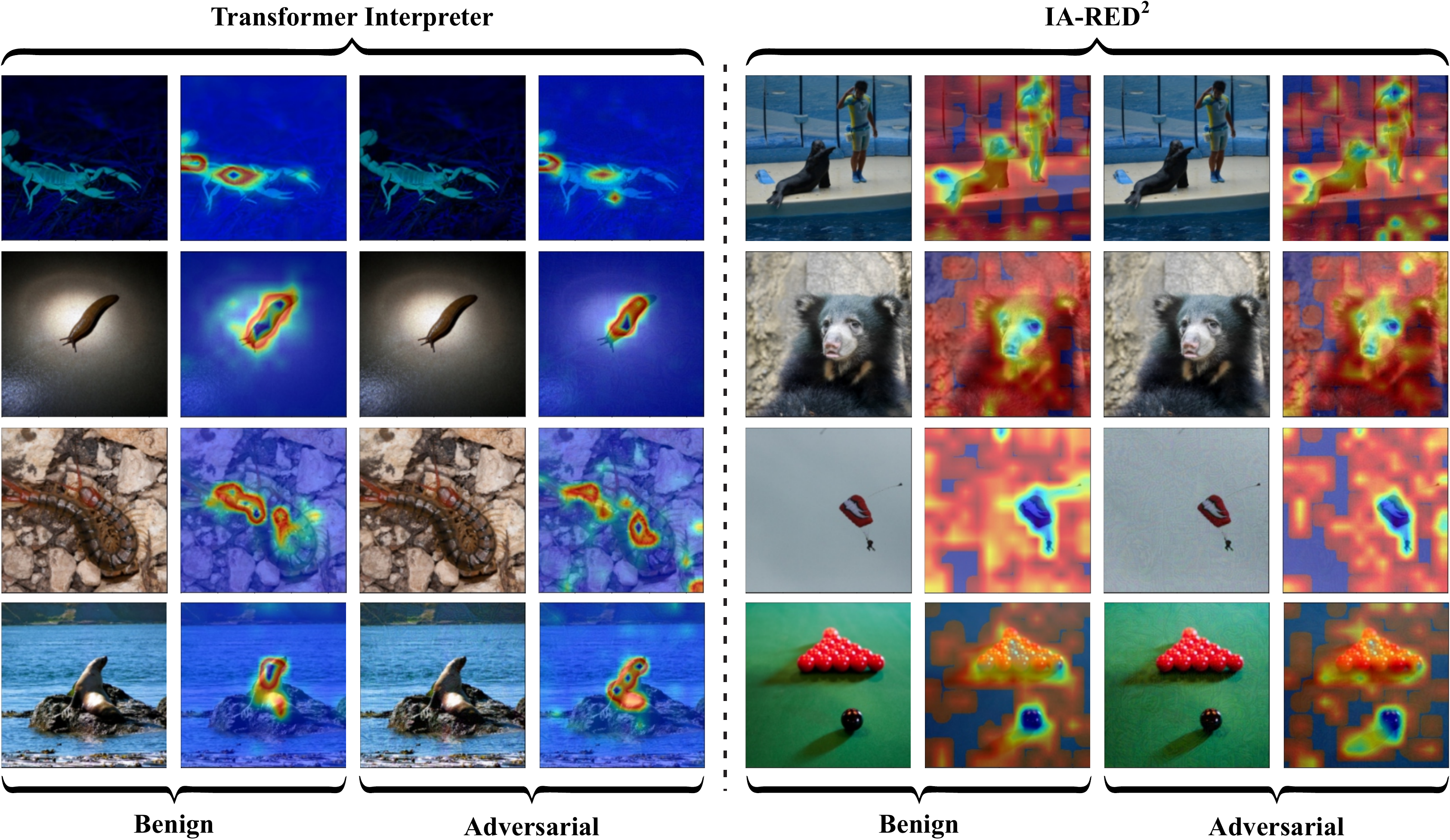}
    \caption{Attribution maps of benign and adversarial samples generated by \ours{} using two interpreters.}
    \label{fig:qualitative_examples}
\end{figure*}

\begin{figure*}
    \centering
    \captionsetup{justification=justified}
    \includegraphics[width=\linewidth]{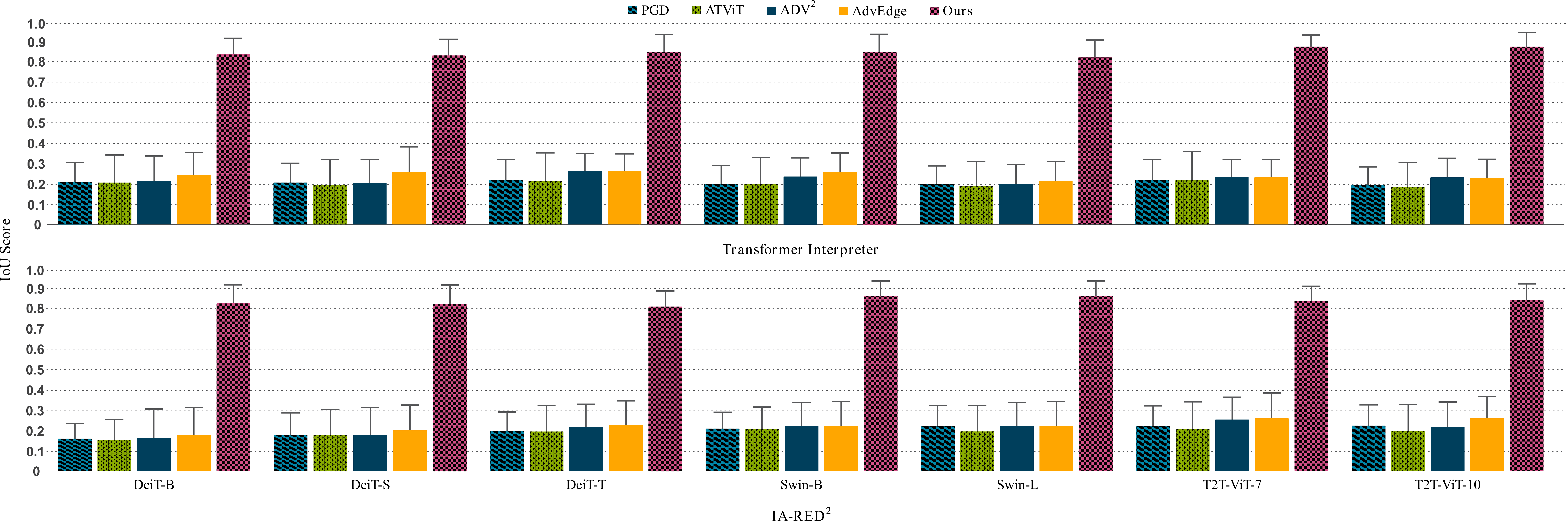}
    \caption{White-box scenario: IoU scores of adversarial interpretation maps generated by \ours{} and existing attacks.}
    \label{fig:iou}\vspace{-1em}
\end{figure*}

\subsection{\ours{}: White-box Settings}\label{subsec:results_whitebox}

This section explores the effectiveness of \ours{} against DeiT-B, DeiT-S, DeiT-T, Swin-B, Swin-L, T2T-ViT-7, T2T-ViT-10, ViT-B, and ViT-L, using two popular ViT interpreters: the transformer interpreter and IA-RED$^2$. In addition to comparing \ours{} with existing interpreter-based adversarial methods (ADV$^2$ \cite{zhang2020interpretable} and AdvEdge \cite{abdukhamidov2021advedge}), we also consider non-interpreter-based baselines, PGD \cite{madry2017towards} and ATViT \cite{naseer2022on}. 

\autoref{tab:mc} shows that our proposed approach successfully misleads all tested models with a 100\% success rate and achieved high misclassification confidence scores. For attacks using the transformer interpreter, ADV$^2$ and AdvEdge obtain average misclassification confidences of approximately 0.62 and 0.63 on the DeiT variants, 0.59 and 0.60 on Swin models, 0.57 and 0.60 on T2T-ViT variants, and 0.61 and 0.59 on ViT models. In contrast, \ours{} consistently outperforms them, reaching 0.72, 0.97, 0.93, and 0.97 on DeiT, Swin, T2T-ViT, and ViT models respectively. 
Under the IA-RED$^2$ interpreter,  ADV$^2$ and AdvEdge struggle to exceed a misclassification confidence of about 0.43, while \ours{} achieves a minimum of 0.36 and generally around 0.60 --- 0.90.

{The results of PGD and ATViT attacks are provided outside the interpreter-based comparison columns on \autoref{tab:mc} as they do not use interpreters to craft their perturbations. Including these non-interpreter-based methods as baselines clarifies that using interpretative information during adversarial generation leads to additional improvements. By extending and refining the fundamental concepts behind PGD and ATViT, \ours{} surpasses these non-interpreter-based methods and significantly advances beyond other interpreter-guided attacks.}

Moreover, our findings suggest that the IA-RED$^2$ interpreter is more robust than the transformer interpreter, as evidenced by the generally lower adversarial confidence scores.

To further analyze the impact of \ours{} on model interpretation, we visualize the interpretation maps of benign and adversarial samples in \autoref{fig:qualitative_examples}. These maps appear nearly identical, indicating that \ours{} preserves critical interpretative features while fooling the classifiers. 
The IoU test results presented in \autoref{fig:iou} confirm this observation: \ours{} achieves IoU scores exceeding 0.8 across all tested transformer-based models for both interpreters. 

In contrast, PGD, ATViT, ADV$^2$, and AdvEdge produce significantly lower IoU values, indicating less accurate interpretation maps.

\begin{table*}
\centering
 \caption{Attack transferability of transformer models to generate adversarial samples against each other. The results are reported as attack success rate (misclassification confidence $\pm$ standard deviation).}
\label{tab:asr_transferability}
\resizebox{0.9\linewidth}{!}{%
\begin{small}
\begin{tabular}{c>{\centering\hspace{0pt}}m{0.09\linewidth}>{\centering\hspace{0pt}}m{0.087\linewidth}>{\centering\hspace{0pt}}m{0.088\linewidth}>{\centering\hspace{0pt}}m{0.087\linewidth}>{\centering\hspace{0pt}}m{0.087\linewidth}>{\centering\hspace{0pt}}m{0.087\linewidth}>{\centering\hspace{0pt}}m{0.087\linewidth}>{\centering\hspace{0pt}}m{0.09\linewidth}>{\centering\hspace{0pt}}m{0.087\linewidth}>{\centering\arraybackslash\hspace{0pt}}m{0.087\linewidth}} 
\toprule
\textbf{} & \textbf{Models} & \textbf{DeiT-B} & \textbf{DeiT-S} & \textbf{DeiT-T} & \textbf{Swin-B} & \textbf{Swin-L} & \textbf{T2T-ViT-7} & \textbf{T2T-ViT-10} & \textbf{ViT-B} & \textbf{ViT-L} \\ 
\midrule
\multirow{18}{0.073\linewidth}{\hspace{0pt}\Centering{}\rotatebox[origin=c]{90}{\textbf{Transformer Interpreter}}} & \textbf{DeiT-B} & \cellcolor[rgb]{0.753,0.753,0.753} & 0.71 (0.62$\pm$0.19) & 0.78 (0.65$\pm$0.18) & 0.31 (0.53$\pm$0.19) & 0.20 (0.51$\pm$0.18) & 0.40 (0.55$\pm$0.13) & 0.41 (0.58$\pm$0.16) & 0.40 (0.94$\pm$0.18) & 0.30 (0.94$\pm$0.19) \\ 
 & \textbf{DeiT-S} & 0.64 (0.56$\pm$0.19) & \cellcolor[rgb]{0.753,0.753,0.753} & 0.70 (0.59$\pm$0.18) & 0.28 (0.48$\pm$0.19) & 0.18 (0.46$\pm$0.18) & 0.33 (0.50$\pm$0.13) & 0.36 (0.52$\pm$0.16) & 0.42 (0.94$\pm$0,18) & 0.32 (0.95$\pm$0.17) \\ 
 & \textbf{DeiT-T} & 0.62 (0.52$\pm$0.18) & 0.57 (0.49$\pm$0.19) & \cellcolor[rgb]{0.753,0.753,0.753} & 0.24 (0.42$\pm$0.19) & 0.16 (0.41$\pm$0.18) & 0.32 (0.44$\pm$0.13) & 0.33 (0.46$\pm$0.16) & 0.39 (0.93$\pm$0.18) & 0.28 (0.95$\pm$0.19) \\ 
 & \textbf{Swin-B} & 0.25 (0.45$\pm$0.14) & 0.26 (0.39$\pm$0.10) & 0.28 (0.51$\pm$0.14) & \cellcolor[rgb]{0.753,0.753,0.753} & 0.57 (0.61$\pm$0.18) & 0.37 (0.35$\pm$0.12) & 0.38 (0.38$\pm$0.14) & 0.25 (0.55$\pm$0.18) & 0.23 (0.51$\pm$0.18) \\ 

 & \textbf{Swin-L} & 0.22 (0.46$\pm$0.15) & 0.20 (0.42$\pm$0.15) & 0.25 (0.44$\pm$0.16) & 0.70 (0.71$\pm$0.19) & \cellcolor[rgb]{0.753,0.753,0.753} & 0.31 (0.30$\pm$0.12) & 0.27 (0.35$\pm$0.15) & 0.21 (0.55$\pm$0.20) & 0.20 (0.56$\pm$0.19) \\ 

 & \textbf{T2T-ViT-7} & 0.19 (0.51$\pm$0.18) & 0.20 (0.44$\pm$0.16) & 0.22 (0.46$\pm$0.16) & 0.19 (0.55$\pm$0.19) & 0.20 (0.50$\pm$0.11) & \cellcolor[rgb]{0.753,0.753,0.753} & 0.77 (0.45$\pm$0.17) & 0.15 (0.54$\pm$0.19) & 0.17 (0.50$\pm$0.18) \\ 

 & \textbf{T2T-ViT-10} & 0.20 (0.52$\pm$0.16) & 0.23 (0.48$\pm$0.18) & 0.23 (0.54$\pm$0.18) & 0.26 (0.57$\pm$0.20) & 0.25 (0.55$\pm$0.18) & 0.84 (0.54$\pm$0.15) & \cellcolor[rgb]{0.753,0.753,0.753} & 0.21 (0.57$\pm$0.18) & 0.20 (0.51$\pm$0.17) \\ 

 & \textbf{ViT-B} & 0.33 (0.58$\pm$0.18) & 0.50 (0.62$\pm$0.20) & 0.57 (0.61$\pm$0.19) & 0.25 (0.49$\pm$0.18) & 0.21 (0.45$\pm$0.13) & 0.29 (0.36$\pm$0.16) & 0.36 (0.38$\pm$0.16) & \cellcolor[rgb]{0.753,0.753,0.753} & 0.81 (0.83$\pm$0.10) \\ 

 & \textbf{ViT-L} & 0.36 (0.60$\pm$0.19) & 0.55 (0.70$\pm$0.20) & 0.58 (0.68$\pm$0.19) & 0.30 (0.47$\pm$0.20) & 0.29 (0.48$\pm$0.19) & 0.35 (0.41$\pm$0.15) & 0.31 (0.44$\pm$0.15) & 0.86 (0.84$\pm$0.10) & \cellcolor[rgb]{0.753,0.753,0.753} \\ 
\midrule
\multirow{18}{0.073\linewidth}{\hspace{0pt}\Centering{}\rotatebox[origin=c]{90}{\textbf{IA-RED$^2$}}} & \textbf{DeiT-B} & \cellcolor[rgb]{0.753,0.753,0.753} & 0.73 (0.44$\pm$0.20) & 0.78 (0.49$\pm$0.19) & 0.32 (0.53$\pm$0.20) & 0.20 (0.41$\pm$0.17) & 0.36 (0.34$\pm$0.20) & 0.36 (0.31$\pm$0.16) & 0.44 (0.61$\pm$0.20) & 0.42 (0.62$\pm$0.20) \\ 

 & \textbf{DeiT-S} & 0.65 (0.39$\pm$0.20) & \cellcolor[rgb]{0.753,0.753,0.753} & 0.70 (0.44$\pm$0.19) & 0.29 (0.48$\pm$0.2) & 0.18 (0.37$\pm$0.17) & 0.32 (0.31$\pm$0.2) & 0.32 (0.28$\pm$0.16) & 0.39 (0.61$\pm$0.16) & 0.34 (0.62$\pm$0.19) \\ 

 & \textbf{DeiT-T} & 0.62 (0.39$\pm$0.2) & 0.58 (0.35$\pm$0.2) & \cellcolor[rgb]{0.753,0.753,0.753} & 0.26 (0.42$\pm$0.2) & 0.16 (0.33$\pm$0.17) & 0.28 (0.27$\pm$0.2) & 0.29 (0.25$\pm$0.16) & 0.33 (0.64$\pm$0.16) & 0.30 (0.63$\pm$0,20) \\ 

 & \textbf{Swin-B} & 0.27 (0.41$\pm$0.18) & 0.27 (0.39$\pm$0.16) & 0.29 (0.40$\pm$0.18) & \cellcolor[rgb]{0.753,0.753,0.753} & 0.71 (0.45$\pm$0.18) & 0.30 (0.33$\pm$0.15) & 0.31 (0.30$\pm$0.16) & 0.26 (0.42$\pm$0.18) & 0.24 (0.38$\pm$0.19) \\ 

 & \textbf{Swin-L} & 0.23 (0.36$\pm$0.18) & 0.22 (0.34$\pm$0.19) & 0.23 (0.38$\pm$0.18) & 0.76 (0.47$\pm$0.19) & \cellcolor[rgb]{0.753,0.753,0.753} & 0.30 (0.33$\pm$0.17) & 0.26 (0.30$\pm$0.16) & 0.23 (0.37$\pm$0.19) & 0.22 (0.31$\pm$0.17) \\ 

 & \textbf{T2T-ViT-7} & 0.25 (0.54$\pm$0.19) & 0.35 (0.47$\pm$0.15) & 0.38 (0.45$\pm$0.15) & 0.33 (0.51$\pm$0.19) & 0.26 (0.50$\pm$0.16) & \cellcolor[rgb]{0.753,0.753,0.753} & 0.78 (0.41$\pm$0.17) & 0.28 (0.51$\pm$0.19) & 0.25 (0.53$\pm$0.19) \\ 

 & \textbf{T2T-ViT-10} & 0.20 (0.42$\pm$0.18) & 0.23 (0.41$\pm$0.17) & 0.25 (0.44$\pm$0.19) & 0.23 (0.53$\pm$0.20) & 0.16 (0.50$\pm$0.17) & 0.81 (0.33$\pm$0.14) & \cellcolor[rgb]{0.753,0.753,0.753} & 0.16 (0.53$\pm$0.19) & 0.16 (0.47$\pm$0.19) \\ 

 & \textbf{ViT-B} & 0.29 (0.47$\pm$0.19) & 0.51 (0.50$\pm$0.20) & 0.54 (0.47$\pm$0.20) & 0.26 (0.47$\pm$0.19) & 0.19 (0.46$\pm$0.12) & 0.33 (0.39$\pm$0.18) & 0.28 (0.47$\pm$0.19) & \cellcolor[rgb]{0.753,0.753,0.753} & 0.78 (0.76$\pm$0.12) \\ 

 & \textbf{ViT-L} & 0.32 (0.46$\pm$0.17) & 0.48 (0.46$\pm$0.19) & 0.49 (0.49$\pm$0.19) & 0.26 (0.43$\pm$0.16) & 0.23 (0.37$\pm$0.19) & 0.31 (0.40$\pm$0.20) & 0.30 (0.38$\pm$0.17) & 0.80 (0.75$\pm$0.11) & \cellcolor[rgb]{0.753,0.753,0.753} \\
\bottomrule
\end{tabular}
\end{small}
}
\end{table*}

\begin{figure*}[tb]
    \centering
    \captionsetup{justification=justified}
    \includegraphics[width=\linewidth]{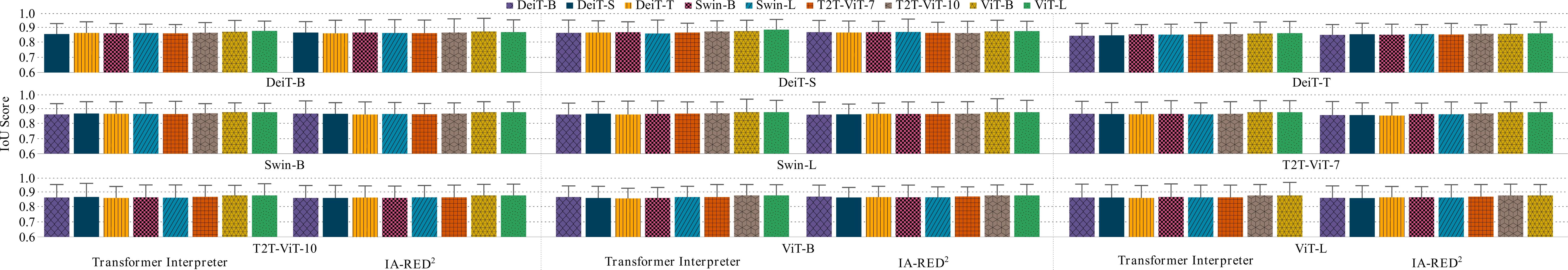}
    \caption{Black-box scenario: IoU scores of adversarial interpretation maps generated by \ours{} using typical transferability and the existing attack.}
    \label{fig:iou_transferability_solid}
\end{figure*}

\subsection{\ours{}: Black-box Settings}\label{subsec:results_blackbox}

Adopting a transferability-based approach (\eg \cite{solans2020poisoning, aivodji2021characterizing}), we
improve transferability using the MGA algorithm. 
We investigate the performance using two settings: \cib{1} typical transferability and \cib{2} improved transferability via MGA. 

\begin{table*}[tb]
\centering
\caption{{Black-box scenario: Comparison of MGA-based \ours{} and Square attack \cite{andriushchenko2020square} in terms of success rate, queries, and confidence. Results for the Square attack are duplicated across both interpreters, as it does not rely on interpreters.}}
\label{tab:asr_mga}
\resizebox{0.9\linewidth}{!}{%
\begin{tabular}{ccc|ccc|ccc} 
\toprule
\multirow{2}{*}{\textbf{Attack}} & \multirow{2}{*}{\textbf{Source Model}} & \multirow{2}{*}{\textbf{Target Model}} & \multicolumn{3}{c|}{\textbf{Transformer Interpreter}}                                                                                       & \multicolumn{3}{c}{\textbf{IA-RED$^2$}}                                                                                                                      \\ 
\cmidrule{4-9}
                                       &  &                                        & \textbf{Success Rate} & \textbf{Avg. Queries} & \begin{tabular}[c]{@{}c@{}}\textbf{Misclassification }\\\textbf{Confidence}\end{tabular} & \textbf{Success Rate} & \textbf{Avg. Queries} & \begin{tabular}[c]{@{}c@{}}\textbf{\textbf{Misclassification}}\\\textbf{\textbf{Confidence}}\end{tabular}  \\ 
\midrule
\multirow{14}{*}{\ours{}} & DeiT-B                                 & \multirow{7}{*}{ViT-B}                 & 1.00                  & 150                   & 0.76$\pm$0.20                                                                                     & 1.00                  & 161                   & 0.54$\pm$0.20                                                                                                       \\ 
& DeiT-S                                 &                                        & 0.90                  & 155                   & 0.72$\pm$0.20                                                                                     & 0.90                  & 158                   & 0.51$\pm$0.20                                                                                                       \\ 
& DeiT-T                                 &                                        & 0.90                  & 180                   & 0.68$\pm$0.20                                                                                     & 0.89                  & 179                   & 0.51$\pm$0.20                                                                                                       \\ 
& Swin-B                                 &                                        & 1.00                  & 120                   & 0.70$\pm$0.14                                                                                     & 0.98                  & 129                   & 0.69$\pm$0.15                                                                                                       \\ 
& Swin-L                                 &                                        & 0.96                  & 128                   & 0.71$\pm$0.14                                                                                     & 0.92                  & 137                  & 0.67$\pm$0.15                                                                                                       \\ 
& T2T-ViT-7                                 &                                        & 0.94                  & 161                   & 0.69$\pm$0.14                                                                                     & 0.93                  & 173                   & 0.70$\pm$0.16                                                                                                       \\ 
& T2T-ViT-10                                 &                                        & 0.94                  & 172                   & 0.66$\pm$0.19                                                                                     & 0.95                  & 184                   & 0.68$\pm$0.19                                                                                                       \\ 
\cmidrule{2-9}
& DeiT-B                                 & \multirow{7}{*}{ViT-L}                 & 0.95                  & 162                   & 0.76$\pm$0.20                                                                                     & 0.97                  & 165                   & 0.57$\pm$0.20                                                                                                       \\ 
& DeiT-S                                 &                                        & 0.87                  & 177                   & 0.70$\pm$0.20                                                                                     & 0.92                  & 185                   & 0.55$\pm$0.20                                                                                                       \\ 
& DeiT-T                                 &                                        & 0.82                  & 188                   & 0.69$\pm$0.20                                                                                     & 0.86                  & 189                   & 0.50$\pm$0.20                                                                                                       \\
& Swin-B                                 &                                        & 0.99                  & 134                   & 0.66$\pm$0.19                                                                                     & 0.96                  & 154                   & 0.67$\pm$0.19                                                                                                       \\ 
& Swin-L                                 &                                        & 0.93                  & 149                   & 0.67$\pm$0.19                                                                                     & 0.94                  & 156                   & 0.67$\pm$0.19                                                                                                       \\ 
& T2T-ViT-7                                 &                                        & 0.90                  & 189                   & 0.70$\pm$0.18                                                                                     & 0.91                  & 183                   & 0.74$\pm$0.19                                                                                                       \\ 
& T2T-ViT-10                                 &                                        & 0.91                  & 195                   & 0.64$\pm$0.19                                                                                     & 0.91                  & 197                   & 0.65$\pm$0.19                                                                                                       \\ 
\midrule
  \multirow{2}{*}{Square} &  & ViT-B & 0.87 & 405 & 0.81$\pm$0.16 & 0.87 & 405 & 0.81$\pm$0.16 \\
 &  & ViT-L & 0.89 & 420 & 0.78$\pm$0.14 & 0.89 & 420 & 0.78$\pm$0.14 \\
\bottomrule
\end{tabular}
}
\end{table*}

\begin{figure*}[tb]
    \centering
    \captionsetup{justification=justified}
    \includegraphics[width=\linewidth]{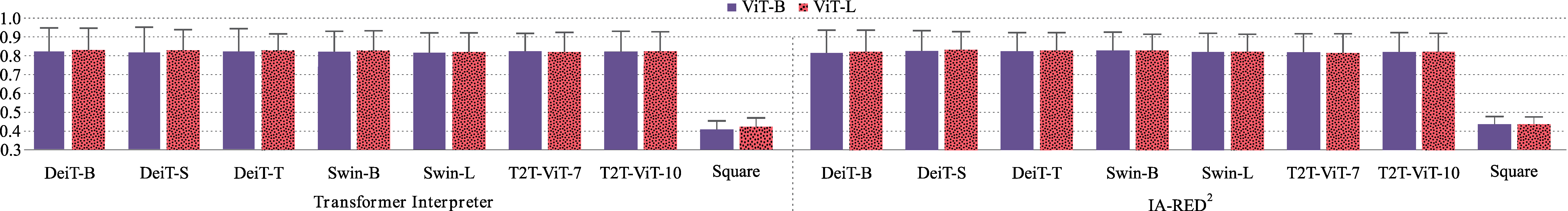}
    \caption{Black-box scenario: IoU scores of adversarial interpretation maps generated by \ours{} using transferability via MGA and Square attack.}
    \label{fig:iou_transferability_mga}\vspace{-1em}
\end{figure*}

\BfPara{Attack Transferability} {We used DeiT, Swin, T2T-ViT, and ViT-based models as a source to generate adversarial samples against each other.} 
\autoref{tab:asr_transferability} shows the attack success rate and misclassification confidence. 
When we used the transformer interpreter in our study, we found different results for each transformer family. For the DeiT family, the highest success rate of the attack was 0.78 (average 0.65 with a variation of $\pm$0.18) and the lowest was 0.16 (average 0.41 with a variation of $\pm$0.18). For the Swin family, the highest was 0.70 (average 0.71 with a variation of $\pm$0.19) and the lowest was 0.20 (average 0.42 with a variation of $\pm$0.15). In the T2T-ViT family, the highest rate was 0.84 (average 0.54 with a variation of $\pm$0.15) and the lowest was 0.15 (average 0.54 with a variation of $\pm$0.19). For the ViT family, the highest was 0.86 (average 0.84 with a variation of $\pm$0.10) and the lowest was 0.24 (average 0.45 with a variation of $\pm$0.13). When we used the IA-RED interpreter, the results were different. For the DeiT family, the highest success rate was 0.78 (average 0.49 with a variation of $\pm$0.19) and the lowest was 0.16 (average 0.33 with a variation of $\pm$0.17). For the Swin family, the highest was 0.76 (average 0.47 with a variation of $\pm$0.19) and the lowest was 0.22 (average 0.31 with a variation of $\pm$0.17). In the T2T-ViT family, the highest rate was 0.81 (average 0.33 with a variation of $\pm$0.14) and the lowest was 0.16 (average 0.47 with a variation of $\pm$0.19). For the ViT family, the highest was 0.80 (average 0.75 with a variation of $\pm$0.11) and the lowest was 0.26 (average 0.43 with a variation of $\pm$0.16).

Furthermore, we investigate the attack transferability
against model interpretation using the IoU test between benign and adversarial attribution maps.
{\autoref{fig:iou_transferability_solid} shows the IoU score of the attack on transformer models with two interpreters. 
As displayed, the performance is significantly high in both interpreters over 0.80.}

\begin{table}[tb]
\centering
\caption{Real-world scenario: attack success rate and misclassification confidence of \ours{} using typical transferability using DeiT-B as source model.}
\label{tab:asr_mc}
\resizebox{\linewidth}{!}{%
\begin{tabular}{lcccc} 
\toprule
\multirow{2}{*}{} & \multicolumn{4}{c}{\textbf{Transformer Interpreter}}                     \\ 
\cmidrule{2-5}
                  & \textbf{ViT-B} & \textbf{SWIN-T} & \textbf{MIT-B} & \textbf{VISION-P}  \\ 

\textbf{Success Rate}      & 0.37           & 0.47            & 0.42           & 0.22               \\ 

\textbf{Misclassification Confidence}       & 0.51           & 0.37            & 0.30           & 0.27               \\
\bottomrule
\end{tabular}
}\vspace*{-2mm}
\end{table}

\begin{table}
\centering
\caption{Real-world scenario: attack success rate, misclassification confidence, and average queries of \ours{} using transferability via MGA and DeiT-B as source model.}
\label{tab:asr_mc_mga}
\resizebox{1\linewidth}{!}{%
\begin{tabular}{lcccc} 
\toprule
\multirow{2}{*}{}                                    & \multicolumn{4}{c}{\textbf{Transformer Interpreter}}                                                      \\ 
\cmidrule{2-5}
                                                     & \textbf{ViT-B}           & \textbf{SWIN-T}          & \textbf{MIT-B}           & \textbf{VISION-P}         \\ 

\textbf{Success Rate}                                         & 0.85                     & 1.00                     & 0.91                     & 0.73                      \\ 

\textbf{Misclassification Confidence}                                          & 0.30                     & 0.26                     & 0.22                     & 0.24                      \\ 

\multicolumn{1}{l}{\textbf{\textbf{Avg. Queries}}} & \multicolumn{1}{c}{205} & \multicolumn{1}{c}{192} & \multicolumn{1}{c}{199} & \multicolumn{1}{c}{218}  \\
\bottomrule
\end{tabular}
}
\vspace*{-3mm}
\end{table}
\BfPara{Improving Transferability via MGA}
{\autoref{tab:asr_mga} shows that \ours{} significantly outperforms the existing Square attack \cite{andriushchenko2020square} in the black-box scenario. \ours{} achieves a 100\% success rate when transferring adversarial samples from DeiT-B to ViT-B under both interpreters, compared to 87\% and 89\% for the Square attack. }
{\ours{} is also more query-efficient, requiring as few as 152 queries for a 100\% success rate on ViT-B, while the Square attack needs 405 queries. Misclassification confidence for \ours{} range from 0.64 to 0.76 with the Transformer Interpreter and 0.50 to 0.74 with IA-RED$^2$, comparable to Square's 0.78 but achieved with higher success and efficiency.}

{Notably, ViT-L is the most robust model against \ours{}, showing slightly lower success rates and requiring more queries. These results demonstrate \ours{}'s effectiveness and efficiency over the Square attack.}

{We evaluate the similarity between adversarial and benign interpretation maps by calculating the IoU score. \autoref{fig:iou_transferability_mga} shows the performance of the proposed \ours{} attack using the MGA algorithm for transferability. Despite the additional noise introduced by the generative algorithm, adversarial interpretation maps remain nearly indistinguishable from their benign counterparts, achieving an IoU of approximately 0.80 in both interpreters. In comparison, the Square attack produces significantly lower IoU scores, around 0.40.}

\begin{table*}[tb]
\centering
\caption{Success rate and average queries of the proposed attack against four defense techniques using different classifiers and interpreters testing on 500 images of ImageNet dataset. The attack is based on black-box settings with MGA transferability.}
\label{tab:defense_result}
\resizebox{0.9\linewidth}{!}{%
\begin{tabular}{ccc|cc|cc|cc|cc} 
\toprule
\multirow{2}{*}{\textbf{Interpreter}} & \multirow{2}{*}{\makecell{\textbf{Source} \\ \textbf{Model}}} & \multirow{2}{*}{\makecell{\textbf{Target} \\ \textbf{Model}}} & \multicolumn{2}{c|}{\textbf{R\&P}} & \multicolumn{2}{c|}{\textbf{Bit-Depth Reduction}} & \multicolumn{2}{c|}{\textbf{Median Smoothing}} & \multicolumn{2}{c}{\textbf{Adversarial Training}} \\ 
\cline{4-11}
 &  &  & \makecell{\textbf{Success} \\ \textbf{Rate}} & \makecell{\textbf{Avg.} \\ \textbf{Queries}} & \makecell{\textbf{Success} \\ \textbf{Rate}} & \makecell{\textbf{Avg.} \\ \textbf{Queries}} & \makecell{\textbf{Success} \\ \textbf{Rate}} & \makecell{\textbf{Avg.} \\ \textbf{Queries}} & \makecell{\textbf{Success} \\ \textbf{Rate}} & \makecell{\textbf{Avg.} \\ \textbf{Queries}} \\ 
\midrule
\multirow{12}{*}{\begin{tabular}[c]{@{}c@{}}Transformer\\Interpreter\end{tabular}} & \multirow{2}{*}{DeiT-B} & ViT-B & 0.95 & 139 & 0.91 & 128 & 0.93 & 121 & 0.97 & 146 \\ 
 &  & ViT-L & 0.90 & 160 & 0.92 & 153 & 0.91 & 138 & 0.94 & 158 \\ 
\cmidrule{2-11}
 & \multirow{2}{*}{DeiT-S} & ViT-B & 0.85 & 147 & 0.88 & 144 & 0.87 & 141 & 0.87 & 195 \\ 
 &  & ViT-L & 0.84 & 188 & 0.86 & 200 & 0.84 & 174 & 0.85 & 207 \\
 \cmidrule{2-11}
 
 & \multirow{2}{*}{Swin-B} & ViT-B & 0.86 & 169 & 0.90 & 153 & 0.90 & 134 & 0.86 & 156 \\ 
 &  & ViT-L & 0.82 & 187 & 0.85 & 158 & 0.87 & 158 & 0.84 & 164 \\ 
 \cmidrule{2-11}
 
 & \multirow{2}{*}{Swin-L} & ViT-B & 0.89 & 172 & 0.89 & 148 & 0.89 & 166 & 0.89 & 210 \\ 
 &  & ViT-L & 0.87 & 207 & 0.87 & 204 & 0.90 & 195 & 0.88 & 211 \\ 

 \cmidrule{2-11}
 & \multirow{2}{*}{T2T-ViT-7} & ViT-B & 0.85 & 165 & 0.87 & 158 & 0.88 & 137 & 0.87 & 158 \\ 
 &  & ViT-L & 0.82 & 185 & 0.88 & 154 & 0.91 & 157 & 0.87 & 167 \\ 

 \cmidrule{2-11}
 & \multirow{2}{*}{T2T-ViT-10} & ViT-B & 0.91 & 180 & 0.94 & 147 & 0.90 & 176 & 0.88 & 201 \\ 
 &  & ViT-L & 0.92 & 205 & 0.92 & 190 & 0.85 & 192 & 0.85 & 206 \\ 

\midrule
\multirow{12}{*}{IA-RED$^2$} & \multirow{2}{*}{DeiT-B} & ViT-B & 0.95 & 110 & 0.97 & 115 & 0.94 & 120 & 0.88 & 150 \\ 
 &  & ViT-L & 0.91 & 144 & 0.94 & 151 & 0.92 & 145 & 0.87 & 184 \\ 
\cmidrule{2-11}
 & \multirow{2}{*}{DeiT-S} & ViT-B & 0.83 & 153 & 0.86 & 163 & 0.83 & 162 & 0.88 & 195 \\ 
 &  & ViT-L & 0.81 & 176 & 0.84 & 182 & 0.80 & 177 & 0.85 & 210 \\

  \cmidrule{2-11}
 & \multirow{2}{*}{Swin-B} & ViT-B & 0.91 & 120 & 0.98 & 124 & 0.98 & 122 & 0.89 & 149 \\ 
 &  & ViT-L & 0.86 & 150 & 0.94 & 152 & 0.90 & 156 & 0.85 & 171 \\ 

 \cmidrule{2-11}
 & \multirow{2}{*}{Swin-L} & ViT-B & 0.86 & 166 & 0.89 & 163 & 0.84 & 169 & 0.83 & 202 \\ 
 &  & ViT-L & 0.80 & 188 & 0.84 & 186 & 0.82 & 189 & 0.84 & 210 \\ 

 \cmidrule{2-11}
 & \multirow{2}{*}{T2T-ViT-7} & ViT-B & 0.89 & 159 & 0.89 & 151 & 0.83 & 144 & 0.89 & 198 \\ 
 &  & ViT-L & 0.91 & 139 & 0.94 & 158 & 0.95 & 154 & 0.88 & 206 \\ 

 \cmidrule{2-11}
 & \multirow{2}{*}{T2T-ViT-10} & ViT-B & 0.88 & 163 & 0.96 & 156 & 0.95 & 164 & 0.90 & 201 \\ 
 &  & ViT-L & 0.82 & 179 & 0.91 & 162 & 0.85 & 175 & 0.86 & 203 \\ 

\bottomrule
\end{tabular}
}
\end{table*}

\begin{figure*}[tb]
    \centering
    \captionsetup{justification=justified}
    \includegraphics[width=\linewidth]{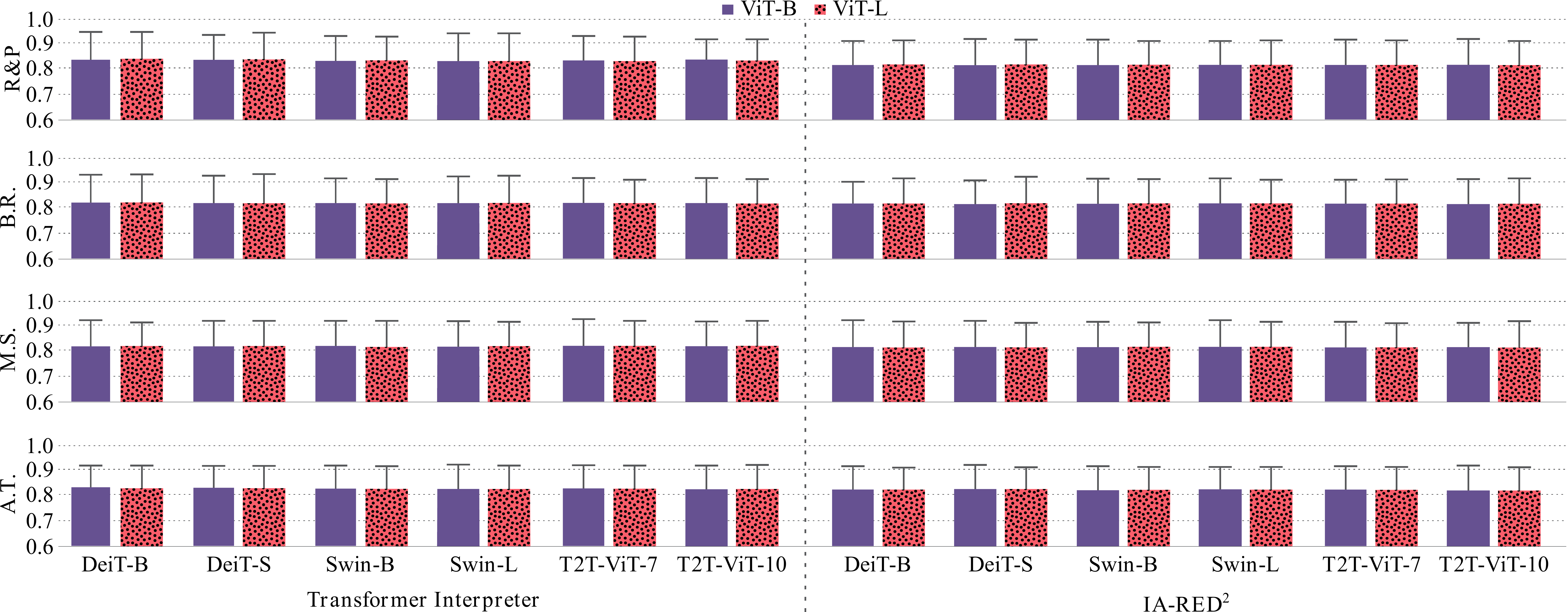}
    \caption{IoU scores of adversarial interpretation maps generated by the proposed attack when defense techniques are applied. MGA algorithm is used to optimize the attack. A.T., M.S. and B.R. stand for Adversarial Training, Median Smoothing, and Bit-Depth Reduction respectively.}
    \label{fig:iou_defense}\vspace{-1em}
\end{figure*}

\subsection{\ours{}: Real-world Black-box Scenario}\label{subsec:results_real_worl_blackbox}

In this experiment, we explore the performance of the proposed attack in real-world scenarios against four different models: ViT-B, SWIN-T, MIT-B, and VISION-P. We conduct the experiment in two transferability settings (see \autoref{subsec:results_blackbox}).
We implement \ours{} using the DeiT-B model as the source model and transformer interpreter.
\autoref{tab:asr_mc} shows the attack success rate and misclassification confidence using a typical transferability setting. 

The result shows that the SWIN-T model is the most susceptible, with an attack success rate of 0.47, while VISION-P is more robust, with the least attack success rate of 0.22 and a misclassification score of 0.27. The ViT-B model shows a high misclassification score of 0.51.

The results show a significant increase in attack success rate when using MGA to enhance transferability, as shown in \autoref{tab:asr_mc_mga}.

For example, the attack success rate increases from 0.22 to 0.73 for the VISION-P model.  
The results show that the average queries are higher for the VISION-P model than for other models.

\subsection{\ours{}: Attacking Defensive ViT Models}\label{subsec:results_defensive}

This section evaluates how MGA-based \ours{} performs when defense techniques are applied in black-box settings. This experiment explores the effectiveness of attack against three well-known pre-processing strategies(\ie R\&P, bit-depth reduction, median smoothing) and adversarial training defense using 500 images from the ImageNet dataset.
For this experiment, we use DeiT, Swin, and T2T-ViT models as source models to generate adversarial samples targeting ViT-B and ViT-L models. 
We investigate the performance of two interpreters: transformer interpreter and IA-RED$^2$. 

Although defense techniques are applied, the attack success rate is still high, as shown in \autoref{tab:defense_result}. For example, the success rate is between 0.82 and 0.95 for the transformer interpreter and between 0.80 and 0.95 in the IA-RED$^2$ interpreter when using R\&P defense. Against adversarial training, the results show that in both interpreters, \ours{} achieves a high success rate ranging between 0.83 and 0.97. Another critical evaluation metric is the number of average queries required to attack the target model in the black-box setting. 
Against all defenses, the average number of queries is between 110 and 210, which is an outstanding result for black-box settings. 

In terms of IoU scores, \autoref{fig:iou_defense} shows that even when a defense technique is applied, our proposed attack still maintains high IoU scores (\ie 0.80 against all scenarios).

\begin{table}[tb]
\centering
\caption{Performance of two types of ensemble detectors composed of two interpreters (\ie transformer interpreter and IA-RED$^2$). The first ensemble detector is based on a 2-channels (\ie 2 interpretation maps) and the second ensemble detector is based on a 3-channels (\ie 3 interpretation maps).}
\arrayrulecolor{black}
\label{tab:two_int_detector}
\resizebox{0.8\linewidth}{!}{%
\begin{tabular}{cc} 
\toprule
\textbf{Detector Type} & \textbf{Detection Success Rate} \\ 
\midrule
2-channel detector & 0.75 \\ 

3-channel detector & 0.80 \\
\bottomrule
\end{tabular}
}
\arrayrulecolor{black}

\end{table}

\subsection{Interpretation-based Adversarial Detection}\label{subsec:results_defensive}
The recent work \cite{zhang2020interpretable} suggests using an ensemble of interpretation models to defend against interpretation-based attacks. 

We test the detectability of the attack based on different interpretations.
Using multiple interpretations of a single input, we build a multiple-interpreter-based detector that checks whether the input is adversarial or benign. 

We generated interpretation maps of adversarial samples via two interpreters for our experiment. For example, adversarial samples and adversarial interpretation maps are generated based on the Transformer interpreter, and extra interpretation maps of those adversarial samples are produced using an IA-RED$^2$ interpreter. We repeated the same process by generating samples based on an IA-RED$^2$ interpreter and applying a Transformer interpreter as a secondary one. 
Since the generated attribution maps are based on single-channel, we stacked single-channel attribution maps from two interpreters to convert them into benign and adversarial two-channel data, respectively. 2,000 benign and 2,000 adversarial samples are produced for the experiment. 

As the dataset size is small, we adopted the pre-trained CNN model EfficientNet-B7 \cite{tan2019efficientnet} to extract feature vectors of a given input and a model called gradient boosting classifier as a final layer instead of the fully-connected layer. This approach is due to the high similarity of benign and adversarial attribution maps and the complexity of the process required to classify the samples. 
Generally, the EfficientNet models have better accuracy and efficiency than the existing CNNs, with a significant reduction in parameter size and FLOPS. 
The gradient boosting classifier consists of several weak learning models that form a stronger predictive model.  
Each attribution map is a one-channel image. To adjust the weights of the model and generate two-channel samples, we replaced the input and output layers of EfficientNet-B7. We used the multiplication of attribution maps extracted from two interpreters as the third channel for the second detector. 
\autoref{tab:two_int_detector} shows the results of the interpretation-based adversarial detector. Even though the dataset is small, the results are promising, which can be seen in the results of the 3-channel detector.

\section{Related Work} \label{sec:related}

\BfPara{Interpretation-guided White-box Attacks} 
Zhang et al. \cite{zhang2020interpretable} conducted the first systematic security analysis for interpretable deep learning systems (IDLSes), demonstrating their vulnerability to adversarial manipulation. They presented ADV$^2$, a new class of attacks that generate adversarial inputs capable of deceiving DNN models and misleading their interpreters. Following this work,
AdvEdge and AdvEdge+ \cite{abdukhamidov2021advedge, 10352932} were proposed to optimize the adversarial attack by adding perturbation to the edges in the regions highlighted by the interpretation map, allowing for more stealthy attacks. Their work has been extended by proposing a query-efficient black-box attack \cite{abdukhamidov2025TR} that stealthily manipulates both predictions and interpretations of deep models without access to model internals. 
In another study, Zhang et al. \cite{zhang2021data} introduced the Interpretation Manipulation Framework (IMF), a data poisoning attack framework that can manipulate the interpretation results of the target inputs as intended by the adversary while preserving the prediction performance. 
Dombrowski et al. \cite{NEURIPS2019_7fea637f} demonstrated that saliency map interpreters (\ie LRP, Grad-CAM) could easily be fooled by incorporating the interpretation results directly into the penalty term of the objective function. 

\BfPara{Interpretation-guided Black-box Attacks} Most existing attacks against IDLSes rely on white-box settings, which limit their practicality in real-world applications.

Zhan \etal \cite{zhan2022towards} introduced a new methodology called Dual Black-box Adversarial Attack (DBAA) that produces adversarial samples to fool the classifier and have comparable interpretations to the benign. They focused on only a single class of interpreters (CAM, Grad-CAM) and CNN-based models. Baniecki and Biecek \cite{Baniecki_Biecek_2022} proposed an algorithm that manipulates SHapley Additive Explanations (SHAP) interpreter based on the perturbation of tabular data. It employs genetic-based data perturbations to control SHAP for a model by minimizing the loss between the manipulated explanation and an arbitrarily selected target. Naseer \etal \cite{naseer2022on} studied improving adversarial transferability in a black-box setting, focusing on ViTs and showing that carefully crafted perturbations can fool models across architectural differences without direct access to their parameters. Our attack strategy builds upon this idea of transferability, leveraging the capacity of robust perturbations to remain effective across various models. proposed SingleADV, a target-specific adversarial attack designed to mislead both predictions and interpretation maps in a class-specific manner. Their method effectively crafts perturbations that suppress visual saliency for the target class while enhancing it for a chosen distractor, exposing vulnerabilities in interpretable deep learning systems. Abdukhamidov et al. \cite{abdukhamidov2025stealthy} proposed SingleADV, a target-specific adversarial attack designed to mislead both predictions and interpretation maps by crafting a universal perturbation for a class-specific category to be misclassified. Their method effectively crafts perturbations that suppress visual saliency for the target class, i.e., maintaining precise and highly relevant interpretations.

\BfPara{Transfer-based Attacks} Aivodji \etal \cite{aivodji2021characterizing} examined the capability of fairwashing attacks by analyzing the fidelity-unfairness trade-off. They demonstrated that fairwashed explanation models generalize beyond the legal group being sued (\ie beyond the data points being explained), suggesting they can rationalize future unfair decisions made on the basis of black-box models using fairwashed explanation models. Fu \etal \cite{10319323} proposed strategies to improve the transferability of adversarial examples across different vision transformers (ViTs) by considering their patch-based inputs and self-attention mechanisms. Zhang \etal \cite{Zhang_2023_CVPR} proposed the Token Gradient Regularization (TGR) method, which reduces the variance of the back-propagated gradient in each internal block of ViTs in a token-wise manner. TGR utilizes the regularized gradient to generate adversarial samples, offering improved performance compared to state-of-the-art transfer-based attacks when attacking both ViTs and CNNs. The transferability of adversarial examples has also been studied in various transfer-based attack methods \cite{10129225, 10375304, 10102830, 10163476}. 

\section{Conclusion} \label{sec:conc}

This work examines the security of IDLSes based on vision transformer models. We present \ours{}, an interpretation-guided attack that generates adversarial inputs to mislead target transformer models and deceive their coupled interpreters.

Through comprehensive experiments, we demonstrate the effectiveness of \ours{} against a range of transformer classifiers and interpretation models in both white-box and black-box settings. We show that \ours{} maintains high transferability to target black-box models, especially when employing MGA to optimize the adversarial samples. 
We also explore the attack's effectiveness against the real-world APIs of four ViT models, ViT-B, SWIN-T, MIT-B, and VISION-P, highlighting the practical implications of our findings. Furthermore, we present the robustness of \ours{} against various defense mechanisms, including random resizing and padding (R\&P), bit-depth reduction, median smoothing, and adversarial training. Although \ours{} demonstrates remarkable success against these defenses, we show that implementing an interpretation-based ensemble detector indicates a promising direction to harden the security of ViT-based IDLSes.

As \ours{} is the first attack targeting ViT models coupled with interpretation models, it paves the way for the development of potentially more powerful adversarial attacks. Our work also serves as a catalyst for researchers to create more effective defenses against attacks similar to \ours{}, fostering a more secure and robust environment for the deployment of ViT-based IDLSes in real-world applications.

\section*{Acknowledgment}
This work was supported by the National Research Foundation of Korea(NRF) grant funded by the Korea government(MSIT)(No. 2021R1A2C1011198), (Institute for Information \& communications Technology Planning \& Evaluation) (IITP) grant funded by the Korea government (MSIT) under the ICT Creative Consilience Program (IITP-2021-2020-0-01821), AI Platform to Fully Adapt and Reflect Privacy-Policy Changes (No. 2022-0-00688), and Convergence security core talent training business (No.2022-0-01199).
{
\balance
\small 
\bibliographystyle{ieee_fullname}
\bibliography{sample-base}
}

\end{document}